S. Fortini[1], G. Querzoli[2], S. Espa[1], A. Cenedese[1]

# Three-dimensional structure of the flow inside the left ventricle of the human heart

*1. Dipartimento Ingegneria Civile, Edile e Ambientale, Sapienza Università di Roma - Via Eudossiana 18, 00184, Roma, Italy*

*2. Dipartimento di Ingegneria Civile Ambientale e Architettura, Università degli studi di Cagliari - Via Marengo 3, 09123, Cagliari, Italy*

e-mail: querzoli@unica.it

## Abstract

The laboratory models of the human heart left ventricle developed in the last decades gave a valuable contribution to the comprehension of the role of the fluid dynamics in the cardiac function and to support the interpretation of the data obtained in vivo. Nevertheless, some questions are still open and new ones stem from the continuous improvements in the diagnostic imaging techniques. Many of these unresolved issues are related to the three-dimensional structure of the left-ventricular flow during the cardiac cycle. In this paper we investigated in detail this aspect using a laboratory model. The ventricle was simulated by a flexible sack varying its volume in time according to a physiologically shaped law. Velocities measured during several cycles on series of parallel planes, taken from two orthogonal points of view, were combined together in order to reconstruct the phase averaged, three-dimensional velocity field. During the diastole, three main steps are recognized in the evolution of the vortical structures: *i)* straight propagation in the direction of the long axis of a vortex-ring originated from the mitral orifice; *ii)* asymmetric development of the vortex-ring on an inclined plane; *iii)* single vortex formation. The analysis of three-dimensional data gives the experimental evidence of the reorganization of the flow in a single vortex persisting until the end of the diastole. This flow pattern seems to optimize the cardiac function since it directs velocity towards the aortic valve just before the systole and minimizes the fraction of blood residing within the ventricle for more cycles.

## Introduction

The general structure of the intraventricular flow is known to affect the effectiveness of the heart as a pump, and possibly cause thrombo-embolism and hemolysis (Ismeno et al. 1999; Alemu and Bluestein 2007; Sengupta et al. 2012).

Since the early twentieth century, technological advances have had a major impact on diagnostic tools, which in these years have had continuous improvements including 3D echo-Doppler cardiography (Haugen et al. 2000; Coon et al. 2012; Benenstein and Saric 2012) ultrasound imaging velocimetry (Poelma et al. 2011) and time-resolved 3D Magnetic Resonance Imaging (MRI) velocimetry (Wigström et al. 1999; Kilner et al. 2000; Töger et al. 2012). Neverthless, our level of understanding of the heart function still represents a limitation to the ability to design strategies for proper diagnosis and effective treatment of cardiac dysfunctions (Vasan and Levy 2000; Mandinov et al. 2000). Thus, the major challenge in solving the problem of dysfunctions in the cardiovascular system arises mainly from the need of a deeper understanding of the basic mechanisms governing the function of the system itself. Improvement in the description of the phenomena based on *in vitro* modelling of the physiological processes is useful for supporting the interpretation of the more and more detailed data acquired *in vivo*.

One of the most important phenomena involved in the left ventricular diastolic flow is the presence of the vortical structures that develop as the filling jet enters through the mitral valve. Some authors focused the interpretation of data obtained from axisymmetric models in terms of vortex dynamics (Steen and Steen 1994; Vierendeels et al. 2002; Baccani et al. 2002) by comparing the space–time map of the axial velocity with clinical echo-Doppler imaging. Other authors focussed on the interaction between vortex generation and leaflet opening in a two-dimensional, experimental model (Romano et al. 2009). The presence of a three-dimensional flow structure has been recognized since the early experimental investigations (Bellhouse 1972; Reul et al. 1981; Wieting and Stripling 1984). These studies showed a transmitral jet, with a leading vortex ring propagating through the left ventricle. In the same years, the phenomenon was numerically simulated using realistically shaped ventricle models (Saber et al. 2001; Lemmon and Yoganathan 2000). The details of the three-dimensional flow during the diastole have also been investigated numerically for an ideally shaped left ventricle corresponding to a healthy child (Domenichini et al. 2005). This study showed that the vortex-ring developing from the mitral valve follows a curved path that turns towards the ventricle wall. Additionally, Pedrizzetti and Domenichini (2005) showed that the intraventricular energy dissipation is significantly affected by the structure of the vortex flow. Other authors used the continuous improvements in MRI and computer processing to develop patient-specific numerical modelling (Schenkel et al. 2009; Doenst et al. 2009).

Experimental ventricle models have been used to obtain time-resolved, 2D velocity fields by means of high-speed cameras and image velocimetry techniques. Most investigators focused on the impact of prosthetic valves on intraventricular flow (Brucker et al. 2002; Cooke et al. 2004; Cenedese et al. 2005, Akutsu et al. 2005, Pierrakos et al. 2005, Querzoli et al. 2010; Vukicevic et al. 2012). The flow was investigated mainly on symmetry planes, assuming two-dimensionality of the velocity field. However, the nature of the flow and the quantities of interest in these studies are intrinsically three-dimensional and these studies do not yield information about the out of plane component of the velocity. Three-dimensional measurements were then necessary to capture all the aspects of the phenomenon.

Simulations *in vitro* have the advantage of being run in controllable and repeatable conditions. Also, they allow for the use of laboratory velocimetry techniques that yield the level of accuracy required for understanding the physical phenomena underlying the complex structure of the flow. In order to study the three-dimensional features of the velocity field, a series of experiments have been run in a flexible, left-ventricle, laboratory model. Vortex dynamics, including the development and propagation of the diastolic vortex-ring and the distribution of viscous shear stresses have been analysed.

## 1. Materials and Methods

The ventricular flow was simulated by means of the laboratory model shown in Fig. 1 and described in detail in Querzoli et al. (2010) and Espa et al. (2012).

A flexible, transparent sack made of silicone rubber (0.7 mm thick) simulated the left ventricle allowing for the optical access.

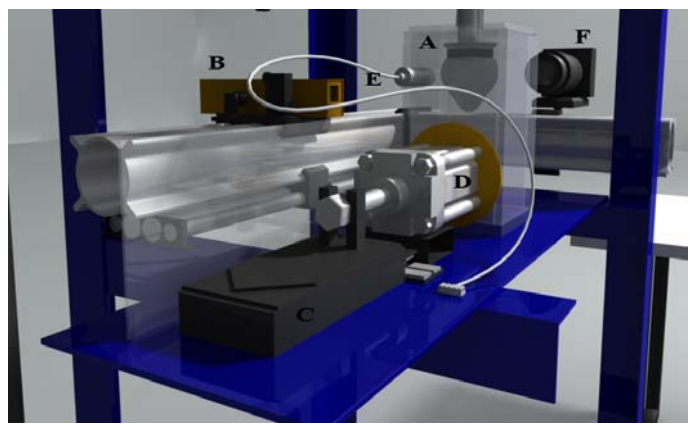

**Fig. 1** Experimental set-up. **A**: Ventricle chamber; **B**: Laser; **C**: Motor; **D**: Piston; **E**: pressure transducer; **F**: high speed camera

The model ventricle was fixed on a circular plate, 56 mm in diameter, connected to a constant-head reservoir by means of two Plexiglas conduits. Along the inlet (mitral) and outlet (aortic) conduits, check-valves were mounted in order to functionally simulate the native heart valves. The inlet (corresponding to the annulus in the real heart) was designed in order to obtain a uniform velocity profile at the mitral orifice. The velocity was verified to be top-hat shaped in a preliminary series of high resolution measurements in the region downstream the inlet (Querzoli et al. 2010). The left-ventricular model was placed in a rectangular tank (A) with transparent, Plexiglas walls. The ventricle volume changed according to the motion of the piston (D), placed on the side of the tank. The piston was driven by a linear motor (C), controlled by means of a speed-feedback servo-control. The motion assigned to the linear motor was tuned to reproduce the volume change acquired *in vivo* by echo-cardiography on a healthy subject. In Fig. 2, the flow rate, q, is plotted as a function of time. The inflow and outflow rates were computed by spatial integration of the velocity over the surfaces corresponding to the inlet and outlet orifice, respectively. The integration was performed by means of a first order scheme.

Velocities were extracted from the phase-averaged data-set obtained as described in the following of this section. The blue line indicates the inflow through the mitral valve during the diastole phase, whereas the red line indicates the outflow through the aortic valve during the systole phase.

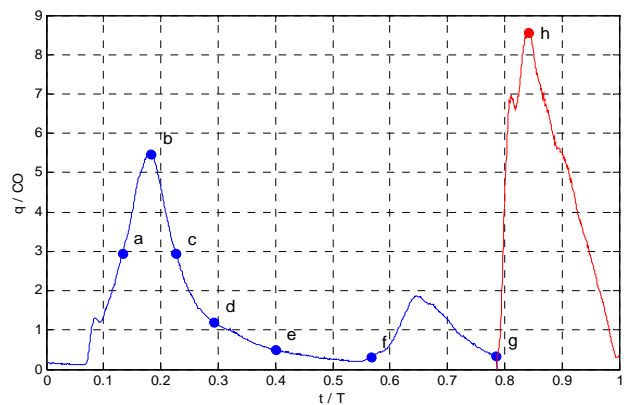

**Fig. 2** Time variation of flow-rate q(t) through the mitral (blue line) and aortic valve (red line) non-dimensionalized by the cardiac output (CO = SV/T, where SV is the stroke volume and T the period of the cardiac cycle). Labelled dots indicate the time instants considered in the next section

The diastole exhibits two peaks: the first (E-wave) corresponds to the ventricle dilation while the second peak is due to the contraction of the left atrium (A-wave). The working fluid inside the ventricle (distilled water) was seeded with neutrally buoyant hollow glass particles with an average diameter of about 30 µm and a density of 1016 Kg/m³. Planes parallel to the long axis of the ventricular model were illuminated by a 12 W, infrared laser (wavelength: 800 nm, light-sheet thickness: 1.5 mm). A triggered high-speed camera (Mikrotron EoSens MC-1362, 250 frames/s, duty cycle: 50%, resolution: 1280x1024 pixels), equipped with a 105 mm - f 2.8, objective, recorded the time evolution of the particle positions at known time intervals for the successive analysis.

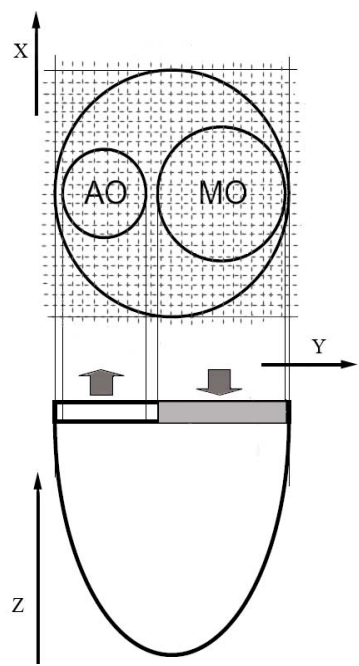

**Fig. 3** Measurement planes (black dashed lines). AO: aortic orifice. MO: mitral orifice.

Data obtained from two sets of planar measurements acquired from two orthogonal views (X-Z and Y-Z, Fig 3) have been used to reconstruct the three-dimensional flow during the ventricle filling. For each view, we acquired 50 cycles on 24 parallel planes distant 2.2 mm from each other. The spatial resolution of the images was, 0.04 mm/pixel. Images of a target immersed in the ventricle were used to check that the ventricle wall was thin enough not to cause meaningful deformation in the images over the investigation regions.

The images of each video recording have been analysed using a Feature Tracking algorithm (Cenedese et al. 2005). This method allowed us to reconstruct the two- dimensional velocity field evolution in a Lagrangian framework. In these experiments, at least 6000 particles have been simultaneously traced during the cardiac cycles.

Interpolation of the velocity vectors over a regular grid for each plane yielded the time evolution of the Eulerian instantaneous velocity field in function of time.

Two-dimensional Eulerian velocity data were phase averaged over the 50 cycles. Finally, the phase averages on the two sets of orthogonal planes were combined together (by linear interpolation) to obtain the phase averaged three-dimensional, three-component, velocity field. The resulting temporal and spatial resolutions of the four-dimensional data set were 4.0 ms and 2.2 mm, respectively. Based on the time and spatial resolution of the video recordings, the accuracy in the measurement of the velocity can be estimated of order of $10^{-4}$ m/s (Miozzi et al. 2008). As a result, a description of the three-dimensional evolution of the intraventricular flow structure during the cardiac cycle was obtained.

Matching the ratio of inertial to viscous effects between the natural heart and the laboratory model requires the equality of Reynolds and Womersley numbers:

$$Re = \frac{U \cdot D}{\nu} \text{ and } Wo = \sqrt{\frac{D^2}{T \cdot \nu}}$$

where D is the maximum diameter of the ventricle, U the peak velocity through the mitral, T the period of the cardiac cycle and ν the kinematic viscosity of the working fluid. The geometrical ratio was 1:1.

Parameters used during the present experiments are: stroke volume 64 ml, T = 6 s, U = 0.145 m/s. The working conditions have been chosen so that the Reynolds and Womersley numbers are within the physiological range: specifically, Re = 8112 and Wo = 22.8.

## 2. Results

The four-dimensional data set was firstly analysed in order to elucidate the role of the coherent structures generated during the diastole. The flow was described in terms of Z-component of the velocity. This quantity catches the main features of the flow such as the filling jet and the vortexes dominating during the diastole, whose axis lay on the X-Y plane. Unlike other quantities commonly used to reveal vortical structures, the vertical velocity field is affected by a low level of noise since its computation does not involve derivatives and, differently from vorticity, it is not sensitive to the shear layer at the walls. Furthermore, data were analyzed to evaluate the spatial distribution of the shear stresses and their relation with the vortical structures. To this aim, the second eigenvalue of the pressure Hessian (Joeng and Hussein 1995), i.e. $\lambda_2$, and the shear stress iso-surfaces have been plotted and discussed.

### 2.1 Vertical velocity

The salient characteristics of the flow can be educed by looking at the regions of ascending (i.e. moving from the ventricle apex to the valves) and descending (i.e. moving towards the apex) fluid. Those regions are identified in Fig. 4 and Fig. 5 by plotting the iso-value surfaces of the Z velocity component corresponding to ± 0.2 U (0.2 is an arbitrary threshold chosen as low as possible but clearly discriminating upwards and downwards motion). The iso-surfaces are presented at the eight time instants indicated with the letters a to h in Fig. 2. The blue surface indicates fluid moving downwards, whereas the red surface indicates fluid moving upwards. Fig. 4 shows four instants during the first filling wave (E-wave). The blue core corresponds to the development of a jet from the mitral orifice that, at the end of the E-wave (t = $t_d$), impinges the ventricular wall near the apex. A vortex ring develops all around the leading edge of the jet and, from the first diastolic peak (t = $t_b$, b), it becomes apparent as far as it induces the upward velocities indicated by the red surface. The core of the vortex ring is located just between the red torus and the blue, downward jet.

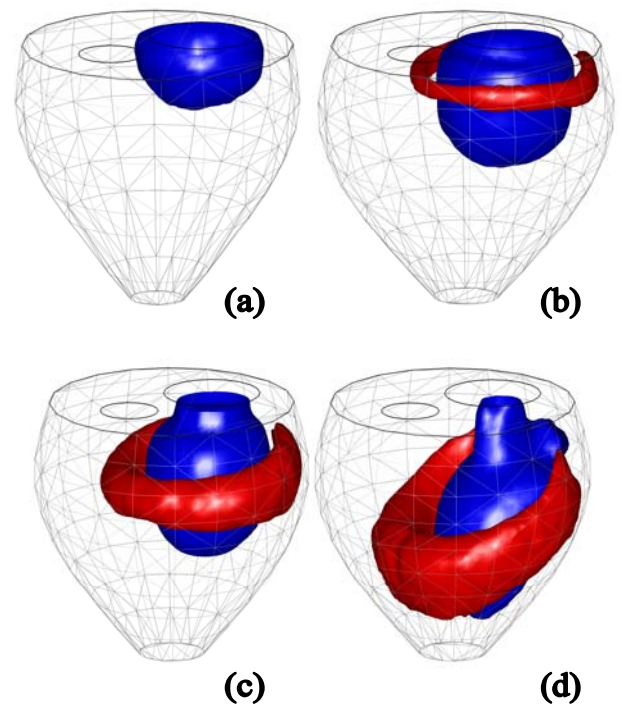

**Fig. 4** Vertical velocity iso-surfaces for values ±0.2 U at times $t_a$ = 0.13 T (a), $t_b$ = 0.18 T (b), $t_c$ = 0.23 T (c), $t_d$ = 0.29 T (d). Red indicates upward velocities, blue indicates downward velocities. Two circles indicate the mitral (larger) and aortic (smaller) orifice

Due to the eccentricity of the position of the mitral orifice with respect to the ventricle axis, the posterior side of the vortex ring (the opposite to the aortic orifice) is closer to the ventricular wall. On that side, the velocity induced by the primary vortex causes a boundary layer to develop at the wall, with the corresponding generation of secondary vorticity. According to the observations of Verzicco and Orlandi (1994), the region of the vortex-ring closest to the wall is characterised by a considerably higher local, stretching rate. Then, its core becomes very thin and the secondary vorticity, produced at the smaller scales, diffuses and annihilates primary vorticity for cross-cancellation. As a consequence, the radius and the intensity of the posterior side of the vortex ring tend to decrease. Moreover, due to the wall effect, it moves slower than its opposite (anterior) side, which is nearly in the centre of the ventricle and increases in radius. Consequently, at the end of the first filling wave ($t = t_d$, d) the vortex ring is oblique (Fig. 5), with the anterior part, close to the apex, which has grown larger. During the time interval between the first and the second filling, the so-called diastasis, the posterior side of the vortex ring vanishes completely, whereas the opposite side continues to grow.

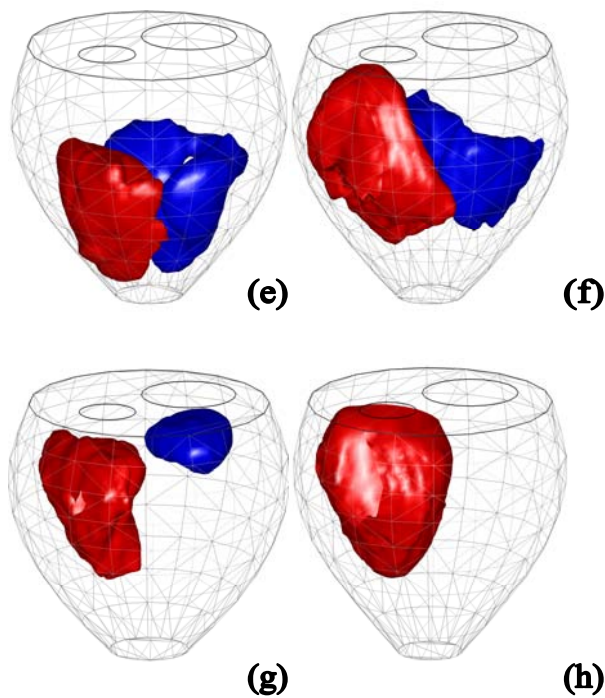

**Fig. 5** Vertical velocity iso-surfaces for values $\pm 0.2$ U at times $t_e = 0.40$ T (e), $t_f = 0.57$ T (f), $t_g = 0.78$ T (g), $t_h = 0.84$ T (h). Red indicates upward velocities, blue indicates downward velocities. Two circles indicate the mitral (larger) and aortic (smaller) orifice, respectively

Correspondingly, the core of the vortex turns into a line, beginning and ending at the ventricle wall, which gradually straightens until it becomes a horizontal line orthogonal to the symmetry plane of the ventricle. As a result, the flow rearranges into a single vortex occupying the whole ventricle. The presence of that large structure is clearly indicated in the plots of figure 5 by the posterior region of downward flow and the anterior region moving upwards at $t = t_e$ (e) and $t = t_f$ (f). The plot at the end of the diastole (Fig. 5, $t = t_g$, g) demonstrates that the second ejection from the mitral, corresponding to the A-wave, does not break the large structure observed at the end of the diastasis, and the main phenomenon remains the (red) region of fluid moving towards the aortic orifice. The above results confirm what inferred from two-components, planar measurements on the symmetry plane (Querzoli et al. 2010, Vukicevic et al. 2012): after the E-wave, the flow reorganizes in a single large structure that prepares the flow to the successive systolic phase. This also agrees with what was observed in four-dimensional MRI flow measurements obtained *in vivo* by Töger et al. (2012). The last plot of Fig. 5 shows that, during the systole, the upwards outflow is the only significant pattern.

## 2.2 Shear Stresses

The vertical velocity plots give information about the general structure of the flow and its evolution during the cardiac cycle.

However, the action of the shear stresses generated by the flow on the blood may play a meaningful role in platelet activation, thrombo-embolism, and hemolysis (Alemu and Bluestein 2007; Nobili et al. 2008).

Though the present spatial resolution (2.2 mm) is not high enough to evaluate effects at the blood cell scale, high-shear regions at the resolved scales correspond to the zones where the above-mentioned phenomena are more likely to take place and then represent a useful description of the large scale flow features.

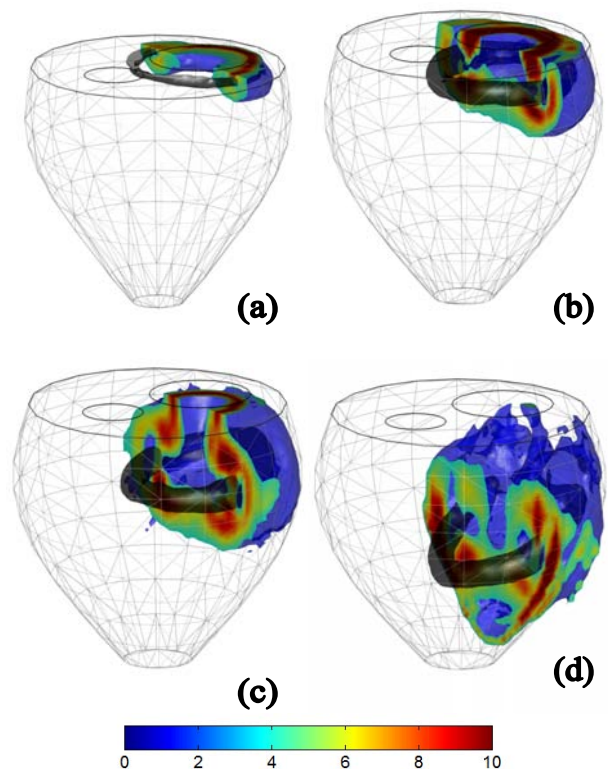

**Fig. 6** Shear Stresses at times $t_a = 0.13$ T (a), $t_b = 0.18$ T (b), $t_c = 0.23$ T (c), $t_d = 0.40$ T (d). Blue surface delimits the region where $\tau_{max} \geq 4$ $\rho U^2$, colours of the slice indicate $\tau_{max}$ ranging from 0 up 10 $\rho U^2$ according to the colour bar. Dark grey $\lambda_2$ iso-surface indicates the vortex-ring core (an arbitrarily negative value was chosen to optimize the identification)

In general, at a given location, the magnitude of the shear stress depends on the orientation of the surface that one considers. However, there is an orientation yielding the maximum shear stress. On that surface, the shear stress can be expressed as (Cenedese et al. 1978; Grigioni et al. 2002, Balducci et al. 2004):

$$\tau_{max} = (\tau_3 - \tau_1) / 2$$

where $\tau_1$ and $\tau_3$ are the minimum and maximum eigenvalue of the stress tensor, respectively. The above maximum value - which is by definition frame independent - is the one discussed in the following. Three-dimensional plots of the maximum shear stresses, $\tau_{max}$, non-dimensionalised by $\rho U^2$ and phase averaged on the whole set of acquired cardiac cycles, are presented at the same time instants as above. Additionally, to give an insight into the relation between the shear stress intensity and the vortical structures characterizing the intraventricular flow, iso-surfaces of the function $\lambda_2$ proposed by Jeong and Hussain (1995) to identify vortex cores have been reported on the same plots. We briefly recall that $\lambda_2$ is the intermediate eigenvalue of the tensor:

$$\Omega^2 + S^2$$

where S and $\Omega$ are the symmetric and antisymmetric part of the velocity gradient tensor, respectively. Vortex cores are typically characterized by high, negative values of $\lambda_2$.

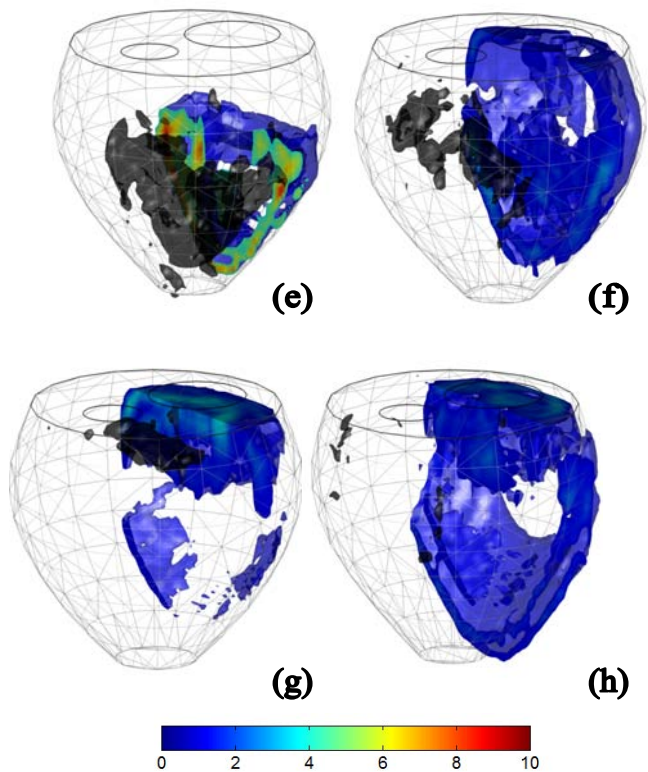

**Fig. 7** Shear Stresses at times $t_e = 0.40$ T (e) $t_f = 0.57$ T (f), $t_g = 0.78$ T (g), $t_h = 0.84$ T (h). Blue surface delimits the region where $\tau_{max} \geq \rho U^2$, colours of the slice indicate $\tau_{max}$ ranging from 0 up 10 $\rho U^2$ according to the lower colour scale. Dark grey $\lambda_2$ iso-surface indicates the vortex-ring core (an arbitrarily negative value was chosen to optimize the identification)

Fig. 6 shows the non-dimensional shear stresses at the same four times of the early filling phase as Fig. 4. During the accelerated ejection ($t = t_a$, a), high shear stress levels are located at the edge of the jet entering the ventricle and just at the inner margin of the vortex-ring core. Comparing Fig. 6a with Fig. 4a, it is worth noticing that $\lambda_2$ criterion can catch the presence of the vortex-ring from its initial stage of development (grey surface). At the first diastolic peak ($t = t_b$, b) the highest shear stresses are still found at the edge of the jet, though a region of elevated values still surround the vortex-ring core. At $t = t_c$ (c) the inflow is decreasing. At the same time, the vortex-ring is fully developed and propagating through the ventricle, relatively free from wall effects. Therefore, the region around the vortex core gives the predominant values, with two peaks just leading and trailing the core.

As the vortex-ring impinges the ventricular wall (Fig. 6, $t = t_d$, d), the high velocities induced by the vortex in the proximity of the wall cause an intense shear. This is the time when the highest shear levels are observed. In the following development of the flow (Fig. 7, e-f), as the flow re-organizes during the diastasis and the second ejection, the shear stress levels are meaningfully lower. Since at this time the vortical motion involves the whole ventricle, the maximum values of the shear stresses are located close to the wall. In particular, the highest values are observed at the posterior ventricular wall, *i.e.* where the jet from the mitral orifice was directed. At the end of the diastasis ($t = t_f$, f), the reorganization of the flow in a single vortex, with a horizontal core orthogonal to the symmetry plane, may be inferred by the approximate alignment of the $\lambda_2$ surface along that direction. The same organization is recognizable also after the second ejection ($t = t_g$, g). However, both the high diffusion undergone by the vortex and the presence of noise, make the eduction of the vortex core from the distribution of $\lambda_2$ not trivial at this phase of the cycle. The vortical structure of the flow vanishes completely during the systole ($t = t_h$, h), and the possibly high shear stresses generated by the outflow are confined within the aortic orifice and in the downstream region, that is out of our measurement field.

## Discussion

The three-dimensional structure of the flow inside the left ventricle has been experimentally analysed and described in terms of vertical velocity, shear stresses and $\lambda_2$ fields during the whole cardiac cycle. It is well known that, during the rapid filling (E-wave), a vortex-ring leads the inflow and propagates through the ventricle (Vierendeels et al. 2002; Cooke et al. 2004). It has been suggested that the vortex-ring plays an important role in optimizing the cardiac function (Pedrizzetti and Domenichini 2005; Dabiri 2009; Querzoli et al. 2010). The continuous enhancement in the details and quality of data obtained *in vivo* suggested, recently, the use of the characteristics of the vortex ring as an index of the efficiency of the left ventricular function (Eriksson et al. 2010; Belohlavek 2012). Most investigators focused on the initial stage of the development of the vortex ring, *i.e.* when it is a well-defined structure, easily detected and measured by Color-Doppler or MRI data (Töger et al. 2012). However, the evolution of the structure of the flow during the entire cardiac cycle is relevant to the efficiency of the pump function of the left ventricle. In this context, the present results give the experimental evidence of the re-organization of the flow during the diastasis in a single, two-dimensional vortex with a horizontal axis, orthogonal to the line connecting the centres of the valve orifices. Consequently, the vortex-ring evolution can be summarized into three main steps:

1. *straight propagation*: of an axi-symmetric vortex-ring parallel to the long axis of the ventricle ($t_a$, $t_b$);
2. *asymmetric development*: the vortex-ring lays on an inclined plane. The posterior side of the ring becomes thinner than the anterior side, which instead increases in radius ($t_c$, $t_d$);
3. *single vortex formation*: the anterior side of the ring gives rise to a single large vortex whose coherence is not broken by the second filling ($t_e$, $t_f$, $t_g$).

Noticeably, phases 1 and 2 are clearly observed in the videos published by Töger *et al.* (2012) as supplemental material to their paper. However, in their investigation they do not go further enough to recognize and discuss the third phase, possibly due to the increasing complexity of the structure of the flow measured *in vivo*.

It has already been recognized that the formation of the vortex-ring stabilizes the filling jet during the E-wave, thus minimizing the production of turbulence (Dabiri and Gharib 2004).

The *single vortex* phase suggests an additional element of optimization in the intraventricular flow. Firstly, it induces blood velocities directed towards the aortic outlet just before the beginning of the systole. Secondly, the presence of a well-organized flow involving a large part of the ventricular volume minimizes the fraction of the blood volume residing inside the left ventricle longer than one cycle.

## Acknowledgements

This work was partially funded by the Ministero dell'Istruzione e della Ricerca Scientifica, PRIN 2009, Project n. 2009J7BL32.